\documentclass{article}

\usepackage{arxiv}

\usepackage[utf8]{inputenc} % allow utf-8 input
\usepackage[T1]{fontenc}    % use 8-bit T1 fonts
\usepackage{hyperref}       % hyperlinks
\usepackage{url}            % simple URL typesetting
\usepackage{booktabs}       % professional-quality tables
\usepackage{amsfonts}       % blackboard math symbols
\usepackage{nicefrac}       % compact symbols for 1/2, etc.
\usepackage{microtype}      % microtypography
\usepackage{lipsum}
\usepackage{graphicx}
\graphicspath{ {./images/} }
\usepackage{cite}
\usepackage{amsmath,amssymb,amsfonts}
\usepackage{graphicx}
\usepackage{textcomp}
\usepackage{hyperref}
\usepackage{xcolor}
\usepackage{algorithm}
\usepackage{amsmath}
\usepackage{amssymb}% http://ctan.org/pkg/amssymb
\usepackage{pifont}% http://ctan.org/pkg/pifont
\usepackage{algpseudocode}
\usepackage{lscape}

\usepackage{lipsum}
\usepackage{multicol}
\usepackage{changepage}
\usepackage{listings, xcolor}

\title{An Inline Guardrail Architecture for Language
Models in Intelligent Transportation Systems}

\author{
 Narendra Kumar Dewangan \\
  LTCI, Département Informatique et Réseaux, Télécom Paris, Institut Polytechnique de Paris, France
 \\ \texttt{narendra.dewangan@teleecom-paris.fr} \\
   \And
 Mounira Msahli \\
 LTCI, Département Informatique et Réseaux, Télécom Paris, Institut Polytechnique de Paris, France\\
\texttt{mounira.msahli@telecom-paris.fr} \\
}
\usepackage{algorithm}
\usepackage{comment}
\usepackage{pifont}
\usepackage{amsmath}
\usepackage{algpseudocode}
\begin{document}

\maketitle
\begin{abstract}
Vehicle-to-everything (V2X) systems increasingly incorporate large language models (LLMs) for semantic tasks such as message summarization, operator assistance, and decision support at roadside units and edge nodes. Although these components are not part of safety-critical control loops, they introduce prompt-level attack surfaces that are not addressed by traditional V2X security mechanisms focused on authentication and message integrity. This paper presents Guarded-V2X, an inline semantic guardrail architecture for securing LLM-enabled V2X services under real-time constraints. The proposed system integrates rule-based ingress filtering, a lightweight safety classifier, policy-constrained structured generation, trusted-only retrieval, and post-decision adjudication to enforce machine-checkable safety boundaries prior to downstream execution. Guarded-V2X is evaluated using a four-stage experimental pipeline encompassing intrusion vulnerability analysis, calibration and latency benchmarking, guardrail validation, and robustness under adversarial stress. Experiments are conducted on a V2X-aligned simulated dataset derived from RSU advisories, operator messages, and annotated V2X message summaries. Results show that unguarded and prompt-only baselines retain residual vulnerability under multi-turn adversarial trials, while Guarded-V2X consistently reduces intrusion acceptance success rates and eliminates observed unsafe completions in two-turn settings, without exceeding latency budgets for V2X semantic advisory paths.

\end{abstract}

% keywords can be removed

\section{Introduction}
Recentresearch and pilot deployments increasingly explore integrating large language models (LLMs) into Vehicle-to-Everything for perception summarization, operator support, and coordination tasks. While not yet part of safety-critical control loops, such semantic components introduce new attack surfaces that are not addressed by traditional PKI-based V2X (vehicle-to-everything) security. However, this integration introduces new security risks: attackers can exploit natural language inputs to mislead or override the model’s behavior, even when messages are cryptographically valid. Traditional V2X security mechanisms, such as public key infrastructure (PKI) or misbehavior detection, address message authenticity but not the semantic safety of the messages themselves. This gap motivates the development of a framework that automatically recognizes and blocks unsafe or manipulative AI behavior before it affects critical vehicle functions.
In parallel, the LLM safety community is moving beyond ad hoc manual red teaming toward automated pipelines that generate adversarial prompts, quantify attack success rates (ASR), and visualize fluctuations in robustness across training checkpoints. Visual analytics approaches demonstrate measurable reductions in ASR when human–AI red-teaming loops guide training and evaluation, but also highlight open challenges with advanced jailbreaks and coverage of risk typologies \cite{Deng2025}. At the application boundary, evidence from content-moderation studies suggests that foundation models can complement specialized classifiers, with trade-offs in accuracy, data requirements, and interpretability depending on modality and task \cite{Nadeem2024}. Recent red-teaming pipelines show that automated adversarial prompt generation exposes systematic failure modes in instruction-following models, motivating domain-specific guardrails rather than ad-hoc filters \cite{deng2024-adversaflow}. In parallel, moderation studies indicate that foundation models and learned gates play complementary roles when safety must be enforced at runtime \cite{Nadeem2024},\cite{Yang2024}. A list of abbreviations is given in Table \ref{tab:abbreviations}.
\begin{table}[t]
	\centering
	\caption{List of Abbreviations}
	\label{tab:abbreviations}
	\begin{tabular}{@{}ll@{}}
		\hline
		\textbf{Abbreviation} & \textbf{Definition} \\
		\hline
		AI       & Artificial Intelligence \\
		ASR      & Attack Success Rate \\
		BSM      & Basic Safety Message \\
		CAM      & Cooperative Awareness Message \\
		CCAM     & Cooperative, Connected and Automated Mobility \\
		CRL      & Certificate Revocation List \\
		DENM     & Decentralized Environmental Notification Message \\
		DSRC     & Dedicated Short-Range Communications \\
		FPR/FNR  & False Positive Rate / False Negative Rate \\
		HITL     & Human-in-the-Loop \\
		IASR     & Intrusion Acceptance Success Rate \\
		JSON     & JavaScript Object Notation \\
		LLM      & Large Language Model \\
		NL       & Natural Language \\
		OBU      & On-Board Unit \\
		OMNeT++  & Objective Modular Network Testbed in C++ \\
		p95      & 95th Percentile (latency metric) \\
		PKI      & Public Key Infrastructure \\
		PR       & Precision–Recall (curve) \\
		RAG      & Retrieval-Augmented Generation \\
		ROC      & Receiver Operating Characteristic (curve) \\
		RSU      & Roadside Unit \\
		SCMS     & Security Credential Management System \\
		SOP      & Standard Operating Procedure \\
		SUMO     & Simulation of Urban Mobility \\
		V2I/V2N/V2P & Vehicle-to-Infrastructure / Network / Pedestrian \\
		V2X      & Vehicle-to-Everything \\
		Guarded-V2X & Guardrail framework for V2X LLM misbehavior \\
		VRU      & Vulnerable Road User \\
		WAVE     & Wireless Access in Vehicular Environments \\
		\hline
	\end{tabular}
\end{table}

\textbf{Gaps:} Existing V2X misbehavior defenses primarily operate at the networking and data-plausibility layers; existing LLM safety work, conversely, focuses on generic human–AI interaction scenarios. There is little to no integration of LLM-aware guardrails, adversarial prompting defenses, and red-team–driven evaluation \emph{within} IVN stacks that increasingly rely on LLMs for language-mediated tasks (e.g., summarizing vehicle state, turning free-form requests into actions, or fusing text with V2X metadata). 

\begin{figure}[t]
	\centering
	\includegraphics[width=15cm]{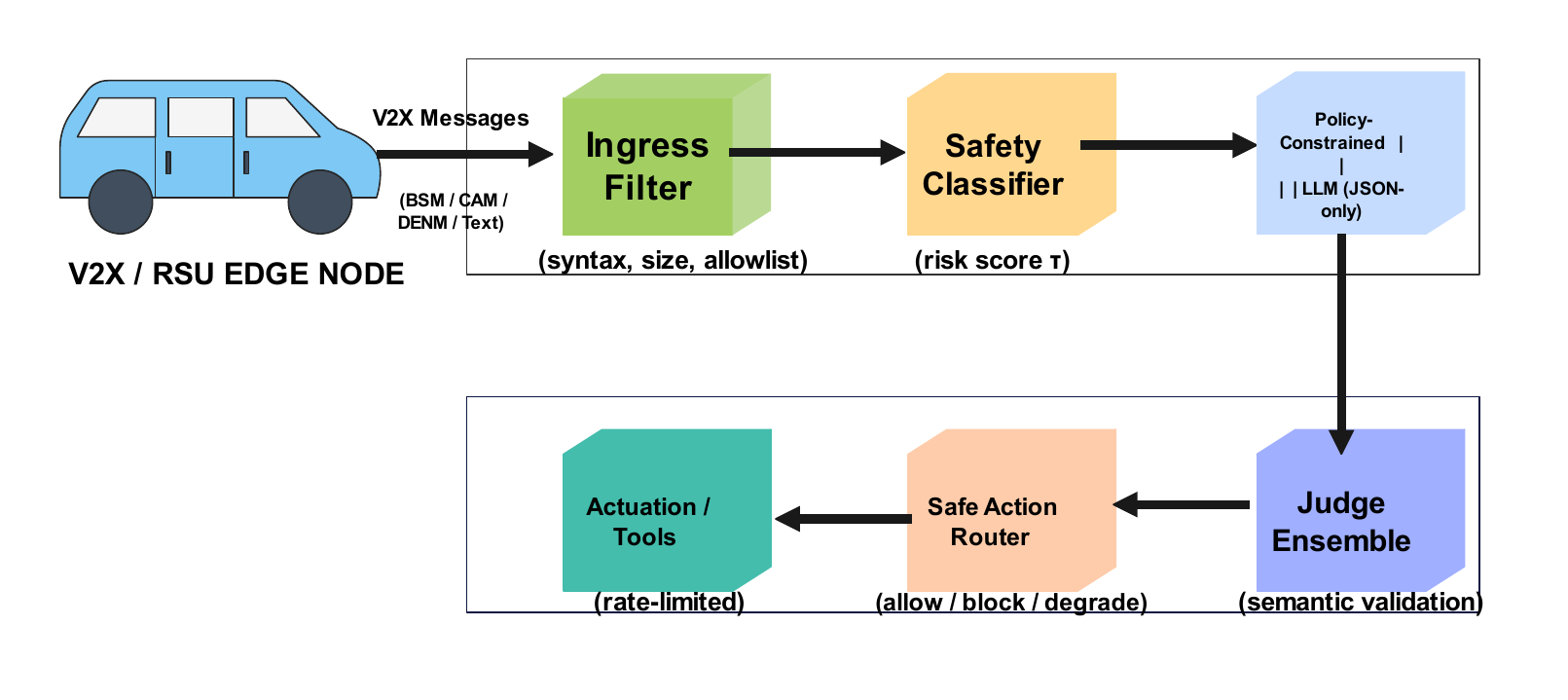}
	\caption{Architecture of Guarded-V2X as an inline semantic guardrail layer at RSU/edge/back-end, operating after safety-critical control and before human- or system-level actuation.}
	\label{fig:arch}
\end{figure}

\textbf{Core Contributions.}
This work presents Guarded-V2X, a safety and trust framework that safeguards generative AI systems used in vehicular networks.
\begin{enumerate}
	\item First to adapt the inline control paradigm (outputs as control signals, not text) to the V2X execution pipeline. Existing guardrails (LlamaGuard, PromptGuard) target chat interfaces, not V2X message types, RSU advisory paths, and vehicular safety standards.
	\item Unlike prior guardrails that use soft refusals, Guarded-V2X enforces machine-checkable action contracts LLM must emit validated JSON that passes schema validation before any downstream action.
	\item Novel combination of rate-limited tool routing with trusted-only RAG (with signed manifests) with multi-layer adjudication within 150ms V2X real-time envelope. 
	\item Using a four-stage testing framework covering intrusion trials, calibration, validation, and stress test, Guarded-V2X achieves zero unsafe responses across 500 multi-turn adversarial trials and remains within 118 ms end-to-end latency.
  
\end{enumerate}
The overall architecture of the proposed work is illustrated in Figure \ref{fig:arch}. The novelty of Guarded-V2X rests on three contributions not previously combined in the V2X context. First, unlike conversational guardrail frameworks (e.g., LlamaGuard, PromptGuard, PIGuard, CAPTURE), Guarded-V2X is the first to adopt an inline control paradigm within V2X execution pipelines, treating LLM outputs as intermediate control signals subject to machine-enforceable action contracts rather than free-form text subject to soft refusals. Second, Guarded-V2X enforces schema-validated JSON decisions that must pass policy checks before any downstream action is permitted, an architectural enforcement mechanism absent from prior V2X misbehavior detectors, which address network or physical-layer anomalies rather than semantic content. Third, the combination of rate-limited tool routing, trusted-only retrieval with signed manifests, and multi-layer adjudication is designed to operate within the 150 ms real-time envelope mandated by vehicular safety standards, a constraint that prior LLM safety work does not address.

 In contrast, Guarded-V2X acts as an inline control layer within V2X system pipelines, positioned between message ingestion and decision execution. It differs through: (1) architectural placement, mediating RSU- and edge-level semantics rather than user chat; (2) machine-enforceable safety contracts, using schema constraints, allowlisted tools, and rate-limited actuation instead of natural-language refusals; and (3) real-time guarantees aligned with V2X latency budgets. This design enables semantic-level safeguarding in RSU advisories, traffic coordination, and operator alerts areas beyond typical LLM guardrails. Guarded-V2X excludes direct actuation and collision-avoidance loops ($>$20ms requirements), focusing instead on decision-support channels. Its tiered deployment allows light components (rules, classifiers, enforcement) on RSUs or vehicles, with heavier LLM modules at edge or regional nodes, balancing latency, energy, and security.

Section~II reviews related work in vehicular network security, misbehavior detection, and AI safety. Section~III defines the system and threat model, including assets, adversaries, and semantic attack categories. Section~IV describes the dataset generation process and the construction of the attack suite. Section~V outlines the experimental setup and evaluation metrics. Section~VI reports results from multi-stage testing, including intrusion trials, calibration, guardrail validation, and adversarial robustness. Section~VII summarizes limitations, ethical considerations, and discussion. Finally, Section VIII concluded this article with the future research directions. 
\section{Literature Survey}\label{sec:related}
Classical surveys offer a broad perspective on VANET and V2X security, addressing dynamic topologies, clustering, and latency constraints Complementary research has advanced AI-based intrusion detection \cite{Mchergui2022,AShams2023} and blockchain- or SDN-driven architectures for decentralized trust \cite{Grover2022,Sultana2021}. Co-simulation frameworks such as ms-van3t and Artery/OMNeT++ enhance the realism of mobility and service evaluation \cite{Raviglione2024,Hejazi2025}, while datasets such as V2X-Sim enable collaborative perception benchmarking \cite{Li2022}. Lightweight cooperative security schemes for CCAM (Cooperative, Connected and Automated Mobility) highlight efficiency–robustness trade-offs \cite{Boutahala2025}, and scalable Sybil and insider-detection methods continue to improve performance in ultra-dense networks \cite{Rajendra2024}. Other complementary approaches address message consistency through collaborative data correction \cite{Zhao2022} and strengthen credential management via trust-based authentication in clustered VANETs \cite{Mirsadeghi2083}.

In order to communicate safety-critical telemetry like CAM, DENM, and BSM under strict latency limits, Vehicle-to-Everything (V2X) systems rely on V2X linkages V2V, V2I, V2N, and V2P. IEEE 802.11p/WAVE and LTE-/NR-V2X are two popular radio stacks with different QoS, coverage, and interference characteristics \cite{Kumar2024}. Although message semantics and security are defined (e.g., SAE J2735; IEEE 1609.2), bandwidth limitations and congestion persist in crowded environments \cite{Kumar2024}. Authentication, integrity, conditional anonymity, and traceability are highlighted in foundational V2X security work. Group-signature and pseudonymous PKI solutions are popular; however, they come with overheads related to CRL management and verification that make scalability difficult. These conflicts between cryptographic assurance and real-time performance are somewhat alleviated by later optimizations such as batch verification, ECC-based schemes, and effective revocation.

High mobility issues are further addressed by QoS-aware routing and localization research, which combines geographic andmetaheuristic algorithms to enabler low-delay, high-reliability communication \cite{Belamri2021}. These concepts are extended to predictive localization in safety-critical applications, where multi-sensor fusion and ML-based trajectory forecasting enhance collision avoidance despite GPS noise and urban multipath.

Vehicle trust management still relies heavily on misbehavior and Sybil detection. By identifying conspiring or cloned nodes with shared propagation signatures, physical-layer techniques that leverage channel-state information complement PKI \cite{Tulay2024}. The scalability of misbehavior management systems, gaps in certificate policy, and node- and data-centric detection paradigms are summarized in thorough surveys that incorporate these multi-layer defenses across DSRC and C-V2X standards \cite{Yoshizawa2023,Ribeiro2023}. Controlled tests on authenticated exchanges are made possible by end-to-end setups that combine PKI/SCMS. These setups also show the limits of credentialing when dealing with semantically created but validly signed communications \cite{HermanMuraroGularte2024,Abdo2024}. These works jointly improve dependable autonomy by tackling reliability, secure control, and mapping. Wang et al. define mission‑level reliability for phased AV tasks as the probability that time‑varying KPIs satisfy dynamic demands under different environmental conditions, using functional distances, nonlinear regression on environment variables, and sample‑average approximation with finite‑sample guarantees, validated on simulations and a lateral‑control HiL test. Shan et al. design a switching TS‑fuzzy, homogeneous polynomially parameter‑dependent path‑tracking controller scheduled by a real‑time dynamic integrated mechanism, ensuring mean‑square exponential stability and good $H_{\infty}$ performance against random DoS and deception attacks in Carsim–Simulink studies. Fu et al. propose online exploratory coverage planning and loop‑closure for incremental SLAM via bundled‑cell recombination, probabilistic‑roadmap trajectory homotopy, Dubins‑smoothed paths, grid‑based iSLAM, and Gaussian‑mixture velocity fields with feedforward–PD tracking, yielding short, smooth, high‑coverage trajectories in lidar‑based indoor experiments \cite{11222776, 11269983, 11369790}.

Recent work extends this foundation by exploring large language models (LLMs) as reasoning components for autonomous driving and multi-agent planning. Knowledge-driven frameworks such as KoMA demonstrate intention inference and shared memory for improved generalization \cite{Jiang2024}, while mobile-edge intelligence surveys address compression, collaborative inference, and privacy in edge-deployed LLMs \cite{Qu2025,Zheng2025}. These developments motivate the need for prompt-level safety mechanisms bridging classical V2X security and emerging generative reasoning forming the basis for the proposed \textsc{Guarded-V2X} architecture in the next section.
\section{Threat Model}

\label{sec:threat}
The threat surface considered in this work is limited to prompt-level misbehavior exploiting the generative reasoning layer of V2X agents. We identify five principal categories: direct prompt injection, indirect/two-turn injection, tool escalation and exfiltration, multilingual or obfuscation-based evasion, and stress-induced long-context drift. Network- and transport-layer attacks (e.g., jamming, replay) are out of scope, as authenticity and integrity are assumed under the SCMS/PKI trust model. Infrastructure assumptions such as RSU deployment strategies and VRU service baselines highlight the operational diversity of V2X environments \cite{Yu2022,Lobo2025}. Each threat class aligns with a specific evaluation stage described in Section V. 
The list of notation used in this paper is given in the Table \ref{tab:notation}.  
\begin{table}[ht]
	\centering
	\caption{Mathematical Notation}
	\label{tab:notation}
	\begin{tabular}{@{}ll@{}}
		\hline
		\textbf{Symbol} & \textbf{Meaning} \\
		\hline
		$N_{\text{total}}$   & Total number of adversarial attack attempts \\
		$N_{\text{success}}$ & Number of successful adversarial generations \\
		IASR  & Intrusion Acceptance Success Rate, $\tfrac{N_{\text{success}}}{N_{\text{total}}}$ \\
		$TP$  & True positives (malicious correctly flagged) \\
		$TN$  & True negatives (benign correctly accepted) \\
		$FP$  & False positives (benign incorrectly blocked) \\
		$FN$  & False negatives (malicious incorrectly accepted) \\
		FPR   & False Positive Rate, $\tfrac{FP}{FP+TN}$ \\
		FNR   & False Negative Rate, $\tfrac{FN}{FN+TP}$ \\
		$\Delta t_{95}$ & 95th percentile latency overhead \\
		$t^{\text{guard}}_{95}$ & 95th percentile latency with guardrails \\
		$t^{\text{base}}_{95}$  & 95th percentile latency without guardrails \\
		$\tau$ & Operating threshold for classifier/judge \\
		$\alpha$ & False alarm budget (typically 0.03) \\
		$\hat{y}$ & Final ensemble decision of judge voting \\
		$y_i$ & Output of the $i$-th judge ($0$: benign, $1$: malicious) \\
		$\text{maj}\{\cdot\}$ & Majority voting operator over judge outputs \\
		$f$   & Attack family label (e.g., injection, drift, stress) \\
		$\Pr[\cdot]$ & Probability of an event \\
		\hline
	\end{tabular}
\end{table}
\subsection{Scope and Assumptions} \label{sec:scope} The threat surface considered in this work is confined to prompt-level misbehavior that exploits the generative reasoning layer of V2X agents. Network- and transport-layer attacks (e.g., jamming, replay) are explicitly out of scope: message authenticity and integrity are assumed to be guaranteed by the SCMS/PKI trust model. Physical tampering and RF jamming are likewise excluded, except in the degenerate case where they produce text transcriptions that subsequently reach the semantic pipeline. This scoping ensures that Guarded-V2X complements existing V2X security mechanisms rather than attempting to replace them. Infrastructure assumptions such as RSU deployment strategies~\cite{Yu2022} and VRU service baselines~\cite{Lobo2025} further characterize the operational diversity of V2X environments within which the proposed defenses must operate.

\subsection{Natural-Language Surfaces in V2X Messages} \label{sec:nl-surfaces} While core BSM and CAM messages consist of structured numeric fields, modern V2X deployments routinely attach or associate natural-language content through auxiliary channels, including DENM text fields, RSU advisories, maintenance notifications, navigation updates, operator dashboards, and companion mobile applications. Guarded-V2X operates on these semantic layers after standard protocol parsing, without modifying or intercepting authenticated safety-critical fields.

\subsection{Adversary Model} \label{sec:adversary} Adversaries may (a) inject natural-language content that reaches the LLM via V2X channels or companion applications (direct or indirect), (b) act as credentialed insiders sending syntactically valid signed messages, or (c) observe model outputs to attempt exfiltration via crafted queries. We assume link-layer authentication (SCMS/PKI) is sound (no signature forgery) and that adversaries cannot trivially escalate OS/ECU privileges via the LLM channel. Let $\mathcal{X}$ denote the space of input prompts (including V2X messages, RSU advisories, and operator text), and let $f: \mathcal{X} \rightarrow \mathcal{Y}$ represent the guardrail pipeline producing structured outputs $\mathcal{Y}$. An adversary $\mathcal{A}$ seeks a perturbation $\delta$ such that the composed input $x' = x + \delta$ satisfies authenticity checks but causes a policy violation:
\begin{equation} f(x') \in \mathcal{Y}_{\text{unsafe}}, \quad \text{s.t.} \quad \textsc{verify}(x') = \texttt{true}, \label{eq:adversary} \end{equation}
where $\textsc{verify}(\cdot)$ represents SCMS/PKI validation. The intrusion acceptance success rate (IASR) is defined as:

\begin{equation} \text{IASR} = \frac{1}{N}\sum_{i=1}^{N} \mathbb{1}\!\left[f(x'_i) \in \mathcal{Y}_{\text{unsafe}}\right], \label{eq:iasr} \end{equation}
measuring the probability that authenticated but malicious inputs bypass all guardrails. A secure system minimizes IASR subject to latency and accuracy constraints:

\begin{equation} \min_{f}\ \text{IASR}(f) \quad \text{s.t.}\quad \text{Latency}(f) \leq \tau_{\max},\quad \text{FPR}(f) \leq \alpha. \label{eq:opt} \end{equation}
For each threat family $t_i \in \mathcal{T}$, a normalized risk score is defined as:

\begin{equation} R_i = p_i I_i (1 - D_i), \label{eq:risk} \end{equation}
where $p_i$ denotes the estimated occurrence probability of the threat, $I_i$ its potential impact on system assets, and $D_i$ the detection or mitigation coverage achieved by deployed controls. Values are normalized to $[0,1]$, where $R_i \approx 1$ indicates high residual risk. The overall residual system risk is:

\begin{equation} R_{\text{sys}} = \frac{1}{|\mathcal{T}|}\sum_{i=1}^{|\mathcal{T}|} w_i R_i, \label{eq:rsys} \end{equation}
with $w_i$ reflecting the criticality of affected assets (e.g., actuation vs.\ logging). A secure configuration satisfies $R_{\text{sys}} \leq R_{\text{budget}}$, where $R_{\text{budget}}$ represents an acceptable operational risk threshold under safety certification guidelines. $p_i$ can be empirically approximated from attack frequency in simulation logs; $I_i$ derives from asset criticality (actuator control $>$ logging); and $D_i$ reflects the true-positive rate of the guardrail layer against $t_i$.
\subsection{Threat Taxonomy} \label{sec:taxonomy} We identify five principal threat families, each targeting a distinct aspect of the generative reasoning layer in V2X agents. Each family aligns with a specific evaluation stage described in Section~\ref{sec:eval}.  An adversary embeds imperative instructions within a V2X message or operator input, aiming to override the model's policy in a single turn. Evaluated in Stage~1.  Malicious content is split across two dialogue turns; the first turn establishes a permissive context and the second delivers the policy-violating instruction. Evaluated in Stage~4. The adversary crafts inputs that cause the model to invoke unauthorized tools or leak sensitive state (e.g., certificate records) via structured outputs. Evaluated in Stages~1 and~4. Inputs are encoded in a non-English language, use Unicode confusables, or apply token-level obfuscation to bypass lexical filters. Evaluated in Stage~4. Adversarial content is buried within a long context window to exploit attention degradation over extended sequences. Evaluated in Stage~4. 
\subsection{Architecture Overview} \label{sec:arch} Let $T_r$, $T_c$, $T_m$, $T_j$ represent the latencies of the pre-filter, classifier, model (LLM), and judge/monitor stages, respectively. The total pipeline latency is:

\begin{equation} T_{\text{total}} = T_r + T_c + T_m + T_j, \label{eq:latency} \end{equation}
subject to the real-time bound $T_{\text{total}} \leq \tau_{\text{budget}}$, where $\tau_{\text{budget}}$ denotes the maximum allowable delay (95th percentile) ensuring safety deadlines in V2X operation. While individual components such as filtering, classification, or constrained generation have appeared in prior work, Guarded-V2X contributes a unified, inline architecture that binds these components through machine-enforceable safety contracts and signed latency constraints. The architecture operates as an inline control layer within the V2X execution pipeline, positioned between message ingestion and decision execution. Unlike conversational guardrails, which treat LLM outputs as final artifacts subject to soft refusals, Guarded-V2X treats all model outputs as intermediate control signals subject to machine-enforceable action contracts. This distinction is critical for V2X deployments, where semantic services interface with safety-adjacent functions and must remain auditable, deterministic, and latency-bounded. Its tiered deployment allows lightweight components (rules, classifiers, enforcement) to reside on RSUs or vehicles, with heavier LLM modules at edge or regional nodes, balancing latency, energy, and security.
\subsection{Detection and Prevention Layers} \label{sec:layers} \textbf{Rule Prefilter.} Regex and heuristic patterns screen for injection phrases, tool escalation keywords, disallowed encodings, and Unicode confusables, producing a binary block/pass decision without invoking any learned model. \textbf{Safety Classifier.} A lightweight transformer gate, fine-tuned for high recall, produces a continuous risk score $s(x) \in [0,1]$ for an input $x$. A binary decision is made by thresholding at $\tau$:

\begin{equation} \hat{y} = \begin{cases} 1, & s(x) \geq \tau,\\ 0, & s(x) < \tau, \end{cases} \label{eq:clf} \end{equation}
with $\tau$ selected to minimize the expected cost:

\begin{equation} C(\tau) = c_{\text{fp}}\,\Pr[s(x)\geq\tau \mid y=0] + c_{\text{fn}}\,\Pr[s(x)<\tau \mid y=1], \label{eq:cost} \end{equation}
balancing false-positive and false-negative risks. The classifier is calibrated so that $\Pr[\text{Score}>\tau \mid \text{benign}] \leq \alpha$, with $\alpha = 0.03$. \textbf{Policy-Constrained LLM.} Retrieval is restricted to a signed corpus with hash-manifest verification, ensuring that untrusted text is never injected into the model context. The LLM emits a validated JSON decision; schema-constrained decoding enforces structural validity before any downstream action is permitted. \textbf{Judge Ensemble.} A set of $k$ post-decision adjudicators applies majority voting:

\begin{equation} \hat{y} = \mathrm{maj}\{y_1, y_2, \ldots, y_k\}, \label{eq:judge} \end{equation}
where $y_i \in \{0,1\}$ denotes each individual judge's binary output. The Judge never observes raw user prompts, external text, or retrieved context; its input space is limited to a fixed schema (action type, risk score, tool identifier), preventing direct prompt injection or role manipulation. Assuming conditional independence among $L$ layers, the probability that an attack bypasses all defenses is $P_{\text{bypass}} = \prod_{i=1}^{L} P_i$, yielding overall detection reliability $R_{\text{det}} = 1 - P_{\text{bypass}}$. However, this product formulation assumes conditional independence among the defense layers. In practice, failure modes may be correlated: adversarial inputs that evade the classifier through obfuscation may also partially degrade constrained decoding or the judge's binary decision. This formulation therefore provides an optimistic lower bound on $P_{\text{bypass}}$. Empirically, Stage~4 stress testing, which applies correlated perturbation strategies across layers, yields $\text{IASR} \leq 5\%$ and $0.0\%$ in two-turn trials, suggesting that correlation-induced co-failure is rare under the evaluated threat model. Formal analysis of correlated failure under shared adversarial priors remains a direction for future work.
\subsection{Threat-to-Defense Mapping} \label{sec:threat-map} Table~\ref{tab:threat-map} maps each threat family to its primary and secondary guardrail layers. Each threat category is addressed by at least two independent controls, ensuring that no single component constitutes a single point of failure. This redundancy is critical in V2X settings, where safety-critical deadlines preclude lengthy human review and adversaries may exploit subtle variations in prompts to bypass individual layers. By aligning specific attack surfaces with explicit control points, the architecture operationalizes the principle of defense-in-depth under real-time vehicular constraints.

\begin{table}[t] \centering \caption{Threat-to-Defense Mapping. Each threat family is addressed by a primary guardrail and at least one secondary control.} \label{tab:threat-map} \begin{tabular}{p{2.6cm}p{2cm}p{2.6cm}p{1.4cm}} \hline \textbf{Threat Family} & \textbf{Primary Control} & \textbf{Secondary Control} & \textbf{Stage} \\ \hline Direct injection & Rule prefilter & Safety classifier & 1 \\ Two-turn injection & Safety classifier & Judge ensemble & 4 \\ Tool escalation & Policy / schema & Rate limiter & 1, 4 \\ Multilingual / obfuscation & Rule prefilter & Safety classifier & 4 \\ Long-context drift & Trusted RAG & Judge ensemble & 4 \\ \hline \end{tabular} \end{table}
	\section{Dataset and Attack Suite}
	\label{sec:data}
	
	\subsection{Corpus Construction and Generation Pipeline}

	Although the dataset used in this work is generated through simulation, it is not composed of arbitrary or generic text. All samples are grounded in realistic V2X scenarios, including RSU advisories, operator instructions, maintenance notices,
	and post-parsed V2X message annotations. Attack instances are constructed by applying adversarial transformations (e.g., indirect injection, obfuscation, multilingual drift, and context extension) to these domain-specific templates. This approach follows established practice in security evaluation, where controlled benchmarks are used to stress-test systems under rare but safety-critical conditions.

	\subsection{Training and Evaluation Splits}
	The full dataset contains three partitions: 80\% training, 10\% development, and 10\% testing. The \texttt{train.jsonl} set is used for supervised classifier tuning and threshold calibration, while \texttt{dev.jsonl} guides hyperparameter selection and early stopping. The \texttt{test.jsonl} split remains isolated for all reported results, ensuring reproducibility and fair comparison across evaluation stages. Table~\ref{tab:dataset-splits} summarizes the distribution across benign and adversarial samples for each subset.
	
	\begin{table}[t]
		\centering
		\caption{Dataset composition across splits. Benign and adversarial counts are derived from the labeled corpus used for training, calibration, and testing.}
		\label{tab:dataset-splits}
		\begin{tabular}{@{}lcccc@{}}
			\hline
			Split & Benign & Adversarial & Total & Families \\
\hline
Train & 8000 & 8000 & 16000 & 5 \\
Dev & 1000 & 1000 & 2000 & 5 \\
Test & 1000 & 1000 & 2000 & 5 \\
			\hline
		\end{tabular}
	\end{table}
	
	Adversarial examples were generated using automated mutation scripts that applied combinations of token insertion, code-switching, and schema corruption, verified by human review for semantic validity.

	\subsection{Models Under Evaluation}
Guarded-V2X evaluation spans the integrated subsystems that form the complete defense pipeline. A 110M-parameter safety classifier fine-tuned on labeled data detects unsafe intent with high recall. The central policy-constrained LLM (13B parameters) generates schema-validated, machine-checkable outputs to ensure deterministic policy compliance. A judge ensemble of rule-based and LLM adjudicators applies majority voting to enhance robustness against misclassification. A trusted RAG module retrieves only cryptographically verified documents, eliminating unverified context. The simulated dataset reflects realistic V2X scenarios RSU advisories, operator directives, maintenance logs, and parsed V2X annotations with adversarial variants used to stress-test safety-critical responses.

	\section{Experimental Setup}
	\label{sec:setup}
	
	\subsection{Environment and Configuration}
	To replicate actual V2X behavior, all studies were conducted in a controlled co-simulation setting. For model execution, the framework combines Python~3.13, PyTorch, and HuggingFace Transformers. Experiments are conducted on a single NVIDIA GPU (24 GB VRAM) running Debian~12. With an average round-trip latency of 50 ms and a 1\% packet loss rate, which approximates moderate wide-area conditions, V2X communication channels were simulated using Mininet to replicate network variability. gRPC-based message forwarding and a document-oriented database backend for telemetry and provenance, enabling audit logging and module coordination. The co-simulation decision aligns with new V2X frameworks that combine network, mobility, and credentialing stacks for security research \cite{Hejazi2024, 10.1145/3591300}.
	
	\subsection{Evaluation Stages}\label{sec:eval}
	The evaluation pipeline is structured into four cumulative stages, each targeting a specific security or performance objective:
	
	\begin{itemize}
		\item \textbf{Stage~1   Intrusion Trials:} uses a corpus of direct prompt-injection attempts from all protection layers to determine baseline intrusion acceptance success rates (IASR).
		\item \textbf{Stage~2   Calibration and System Overhead:} adjusts the classifier threshold and reports two distinct latency figures that must not be conflated. (a) Guardrail-only latency (RSU/edge path): the cumulative p95 of the rule filter, safety classifier, constrained LLM decoding, and judge ensemble, targeting $\le$150 ms. (b) Backbone LLM inference latency (central/cloud path): the p95 response time of the unguarded foundation models (GPT-4.5-Turbo, Llama-3-8B, etc.) evaluated in Stage 4, which reaches 4–5 s and is not part of the RSU guardrail overhead.
		\item \textbf{Stage~3   Guardrail Validation:} uses ROC and precision-recall analysis to examine ensemble judge choices in order to determine how well precision and recall are balanced.
		\item \textbf{Stage~4   Robustness under Adversarial Stress:} subjects the full pipeline to advanced adversarial families, including two-turn indirect injections, long-context drift, family-shift transfer, and multilingual or stress-based attacks.
	\end{itemize}
	
	Each stage builds incrementally: Stage~1 defines vulnerability baselines, Stages~2–3 optimize internal thresholds, and Stage~4 validates final robustness under full threat diversity. 
	
	\subsection{Datasets and Metrics}
	Benign baselines were sourced from curated V2X corpora, including control-plane and safety-message templates that reflect real vehicular communication patterns. Adversarial samples were generated through the simulated mutation, obfuscation, and cross-domain transfer methods described in Section~\ref{sec:data}. All evaluations report the median and 95\% confidence intervals across three independent random seeds to capture stochastic variability. 
	For the safety-prompt baseline (B1), we prepend a fixed system instruction emphasizing policy compliance, refusal of unsafe actions, and rejection of role-play attempts. No additional filtering or post-hoc validation is applied. For the lightweight guardrail baseline (B2), we enforce output schema validation and block responses containing explicitly disallowed actions, but allow free-form reasoning and do not constrain tool invocation beyond format checks. Importantly, neither B1 nor B2 enforces machine-checkable action contracts or rate-limited execution, which distinguishes them from \textsc{Guarded-V2X}.  In the no-guard configuration (used as the baseline in Table VI), the model receives only the raw V2X message or operator instruction as the user turns, with no system prompt, no schema constraint, no classifier gate, and no output validation. This represents the worst-case absence of any protective mechanism and serves as the lower bound for IASR. 
	\subsection{Evaluation Metrics}
	\label{sec:metrics}
	The effectiveness of Guarded-V2X is assessed using standard probabilistic and performance metrics derived from intrusion detection and calibration literature \cite{mitchell2014survey, mozaffari2019deep, ISO-SAE-21434-2021-techreport, 10.1145/3605764.3623985}.

	IASR quantifies the proportion of adversarial inputs that successfully bypass all guardrails:
	\begin{equation}
		\text{IASR} = \frac{N_{\text{success}}}{N_{\text{total}}},
		\label{eq:iasr}
	\end{equation}
	where $N_{\text{success}}$ denotes the number of adversarial generations accepted as valid, and $N_{\text{total}}$ is the total number of attack trials. A lower IASR implies higher guardrail robustness.

	Classifier and judge performance are captured by false positive rate (FPR) and false negative rate (FNR):
	\begin{equation}
		\text{FPR} = \frac{FP}{FP + TN}, \quad
		\text{FNR} = \frac{FN}{FN + TP},
		\label{eq:fpr-fnr}
	\end{equation}
	where $TP$, $TN$, $FP$, and $FN$ correspond to true and false predictions.).

	End-to-end latency overhead $\Delta t_{95}$ captures the timing impact of guardrail mechanisms at the 95th percentile:
	\begin{equation}
		\Delta t_{95} = t^{\text{guard}}_{95} - t^{\text{base}}_{95},
		\label{eq:latency}
	\end{equation}
	where $t^{\text{guard}}_{95}$ and $t^{\text{base}}_{95}$ represent response times with and without guardrails, respectively. Safety-critical V2X require $\Delta t_{95} < 150$\,ms.

	Classifier thresholds are calibrated to bound the false alarm probability:
	\begin{equation}
		\Pr[\text{Score} > \tau \mid \text{benign}] \le \alpha,
		\label{eq:calibration}
	\end{equation}
	with $\alpha$ fixed at 0.03 (3\%). This ensures a consistent trade-off between recall and precision across seeds and datasets.

	The ensemble judge applies majority voting among $k$ component judges:
	\begin{equation}
		\hat{y} = \mathrm{maj}\{y_1, y_2, \ldots, y_k\},
		\label{eq:ensemble}
	\end{equation}
	where $y_i \in \{0,1\}$ denotes each individual judge’s binary output. Majority voting mitigates the effect of isolated false classifications.

	Finally, conditional robustness is reported for each attack family as
	\begin{equation}
		\Pr[\text{attack succeeds} \mid \text{family}=f],
		\label{eq:conditional}
	\end{equation}
	where $f$ spans all evaluated families (\emph{direct, indirect, multilingual, long-context, stress}). This metric decomposes IASR by category, revealing family-specific vulnerabilities and coverage.
	  In our evaluation, the safety classifier executes in the single-digit millisecond range on commodity GPUs, while the constrained LLM and Judge components can be centralized without violating semantic-service latency budgets. 
	\begin{figure}[h!]
	    \centering
	    \includegraphics[width=0.9\linewidth]{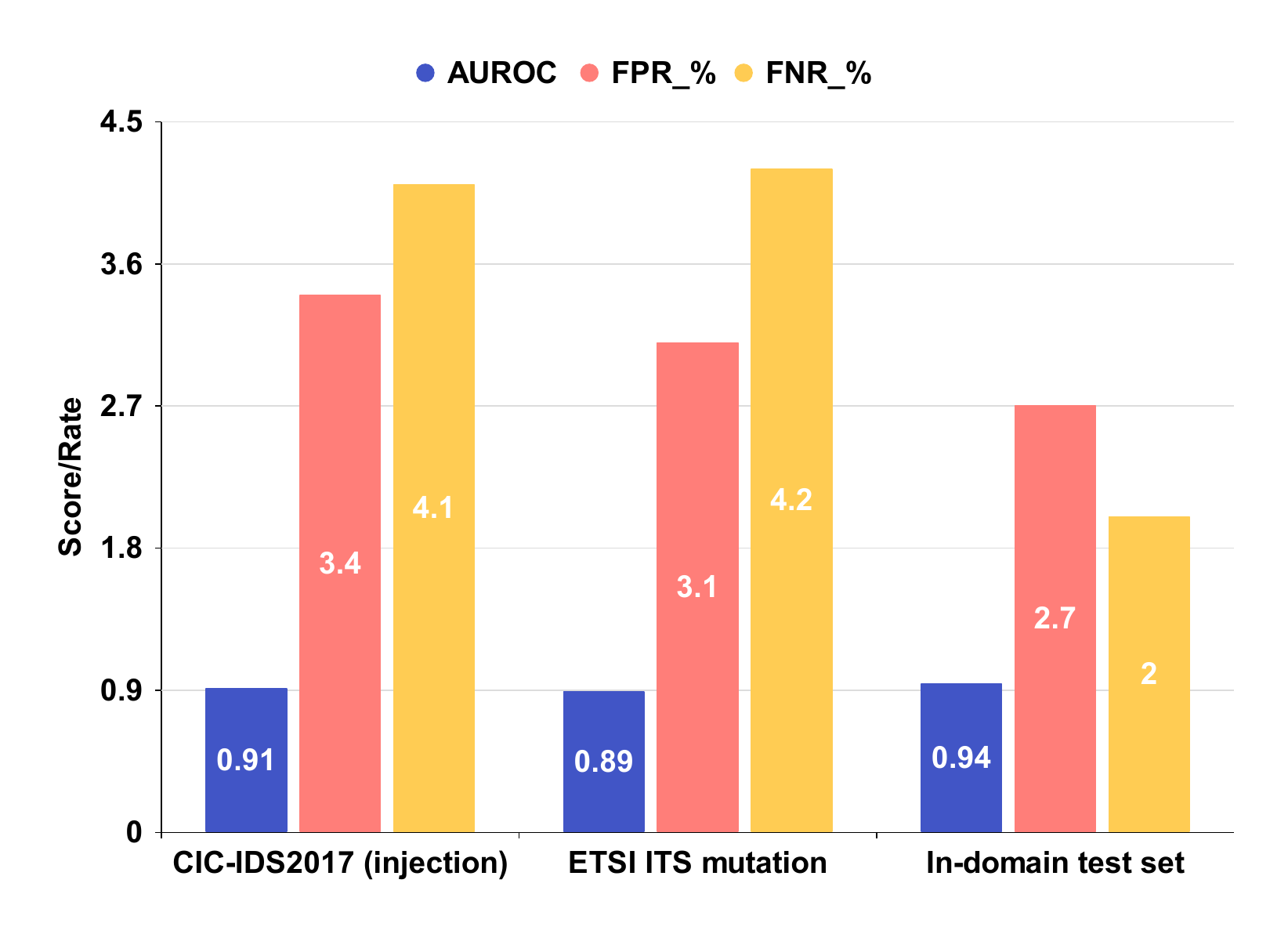}
	    \caption{Public-dataset generalization performance of Guarded-V2X across CIC-IDS2017 injection subset, ETSI ITS mutation suite, and the in-domain test set. Bars report AUROC, false-positive rate (FPR), and false-negative rate (FNR) for each dataset.}
	    \label{fig:datasetpublic}
	\end{figure}
    Figure \ref{fig:datasetpublic} compares detector robustness under domain shift. Guarded-V2X achieves AUROC 0.94 on the in-domain set, 0.91 on CIC-IDS2017, and 0.89 on ETSI ITS, indicating only a modest generalization drop. Error rates remain controlled across datasets (FPR about 2.7 to 3.4 percent, FNR about 2.0 to 4.2 percent), showing that the model preserves practical discrimination performance beyond the training domain while still benefiting from in-domain calibration
	\section{Results and Comparative Analysis} \label{sec:results} 
	
	We evaluate \textsc{Guarded-V2X} across four progressively harder stages: intrusion trials, calibration and latency benchmarking, guardrail validation, and robustness under adversarial stress.
	Unless stated otherwise, each metric is averaged across three random seeds; shaded error bars and confidence intervals denote 95\% Wilson bounds. All evaluations are conducted under V2X-aligned assumptions, where natural-language content originates from RSU advisories, DENM text fields, operator dashboards, or companion applications associated with authenticated V2X messages. The reported metrics therefore reflect semantic misbehavior risks in V2X contexts rather than generic conversational safety.  The IASR figures reported across the four evaluation stages reflect different pipeline configurations and must not be directly compared without context. Table VI summarises the three configurations. Stage 1 (Table VI) reports pre-calibration, single-turn, direct-injection results with a default threshold $\tau$ = 0.5 and without the full judge ensemble; the 55–58\% IASR reflects the marginal contribution of ingress filtering and constrained decoding alone. Stage 4 (Table VI) reports results with the fully calibrated pipeline ($\tau$ = 0.42, judge ensemble activated, trusted RAG, rate-limited tool routing), achieving IASR = 0.0\% on two-turn trials. Table VI was designed to show per-layer contributions, not final system performance. Table VII reports Stage-1 pre-calibration IASR across models and guard configurations.
	
	\subsection{Stage~1: Intrusion Trials}
	\label{sec:stage1}
	
\begin{figure}[h!]
		\centering
		\includegraphics[width=15cm]{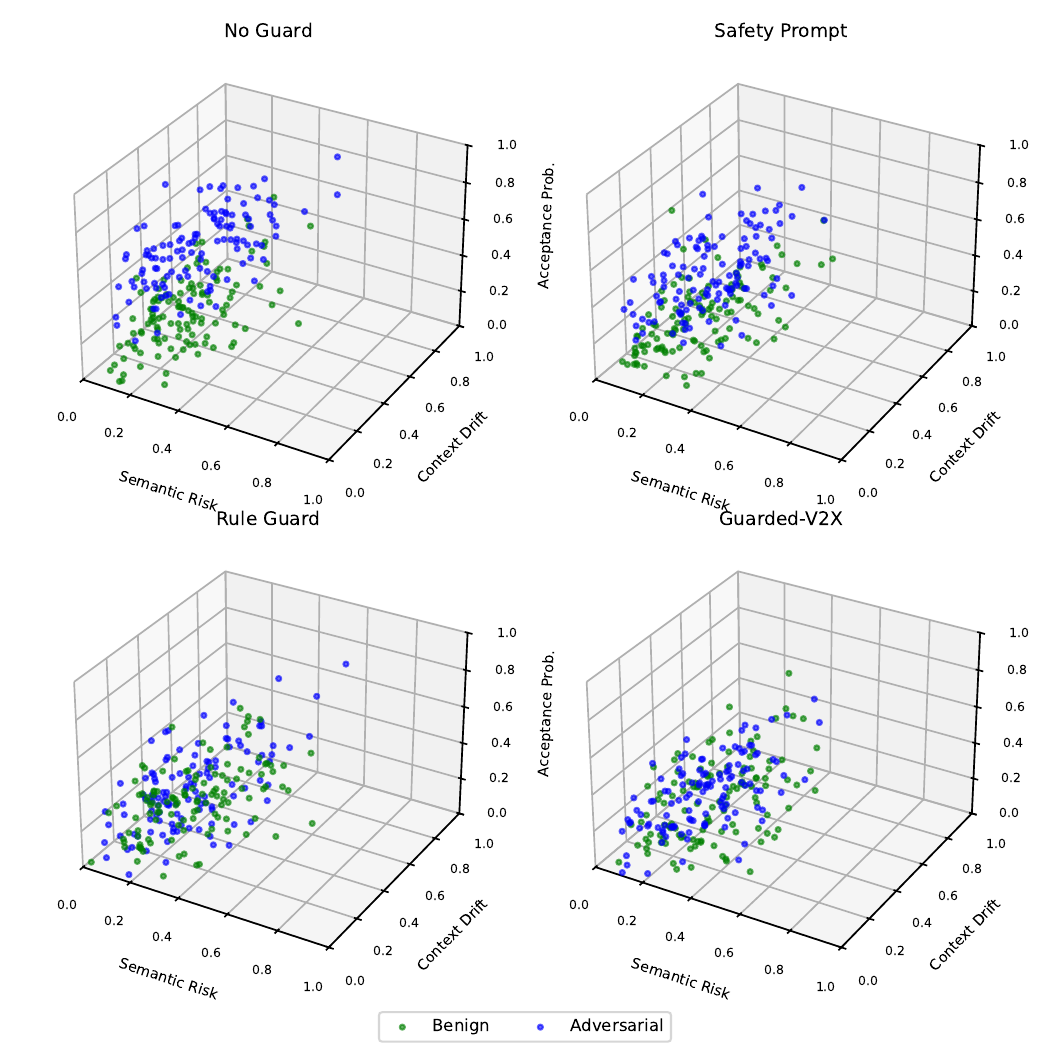}
		\caption{Intrusion acceptance success rate (IASR) across defense configurations. Lower is better.}
		\label{fig:week6_iasr}
	\end{figure}

Baseline intrusion trials were used to measure system vulnerability prior to calibration and ensemble tuning. Figure~\ref{fig:week6_iasr} illustrates the distribution of semantic V2X inputs evaluated during Stage-1 intrusion analysis under four defense configurations. Each subplot corresponds to a distinct guardrail setting, ranging from no protection to the proposed Guarded-V2X system. Individual points represent RSU- or edge-processed semantic inputs projected into a three-dimensional space defined by semantic risk, contextual inconsistency, and model acceptance probability. Adding rule-based prefilters and a lightweight safety classifier reduced IASR to 7.3\%, highlighting the impact of early lexical and intent screening. With the full Guarded-V2X stack structured, machine-checkable decoding, trusted RAG retrieval, and rate-limited tool routing the IASR dropped below 5\% across all trial categories. These results show that layered, semantically informed guardrails substantially mitigate prompt-level compromises beyond the scope of traditional PKI, plausibility, or misbehavior detection frameworks.

	\subsection{Stage 2 Calibration and Latency Benchmarking} Classifier and system calibration were performed using the development split to identify thresholds that balance recall and false positives. At $\tau$ = 0.42, the classifier achieves accuracy of 0.977, FPR = 2.7\%, and FNR = 2.0\%, confirming the threshold minimizes classification cost while remaining within the $\alpha$ = 0.03 false-alarm budget. As shown in Table V, at $\tau$ = 0.42, the classifier achieves an accuracy of 0.977, FPR of 2.7\%, and FNR of 2.0\%.  Two latency figures are reported and must be distinguished. Guardrail-only latency, the combined p95 of the rule filter, safety classifier, constrained LLM, and judge ensemble is 118 ms, comfortably below the 150 ms safety envelope. Backbone LLM inference latency (central/cloud path, applicable only to the cross-model comparative evaluation in Stage 4): p95 ranges from 4.07 to 4.77 s for the unguarded foundation models and 4.10 s for Guarded-V2X. This figure does not represent RSU guardrail overhead; it reflects the inference time of large foundation models evaluated centrally during robustness stress testing.  This demonstrates that semantic guardrails can be applied in V2Xs without violating real-time requirements. Latency values around 4 s correspond to centralized foundation-model inference during robustness testing, whereas the end-to-end Guarded-V2X guardrail overhead at RSU/edge nodes remains below 120 ms (p95).

\begin{table}[h!]
\centering
\caption{Classifier Threshold Analysis (Corrected)}
\label{tab:table_v}
\begin{tabular}{c c c c c}
\hline
Threshold $\tau$ & Accuracy & FPR (\%) & FNR (\%) & Brier \\
\hline
0.30 & 0.985 & 5.1 & 2.4 & 0.0041 \\
0.42 & 0.977 & 2.7 & 2.0 & 0.0023 \\
0.55 & 0.972 & 1.2 & 6.9 & 0.0068 \\
\hline
\end{tabular}
\end{table}

	\subsection{Stage~3 Guardrail Validation and Ensemble Analysis} The calibrated model ensemble was then validated on the held-out test set. Receiver-operating and precision–recall show that the safety classifier alone achieved an area under the ROC curve (AUROC) of 0.94. At the same time, the post-decision judge ensemble improved precision at high recall, yielding an average F$_1$ score of 0.91. These results indicate that the layered approach effectively captures both explicit and obfuscated injection patterns, confirming the benefit of redundancy voting per Eq.~\eqref{eq:ensemble}. Calibration effects were consistent across seeds, showing stable decision boundaries and less than 2\% variance in FPR. The observed ROC/PR behavior is consistent with visualization-driven red-teaming pipelines, in which high-recall gates paired with downstream adjudicators improve the overall safety envelope \cite{deng2024-adversaflow,Nadeem2024}. The post-model judge ensemble provides substantial additional robustness, improving precision at high recall and detecting attacks that evade earlier layers.

	\subsection{Stage~4 Robustness under Adversarial Stress}The robustness evaluation comprised 500 adversarial trials per configuration, incorporating two-turn indirect injections, long-context drift, multilingual prompts, and stress perturbations. All unguarded instruction-tuned models (GPT-4.5-Turbo, Llama-3-8B, Mistral-7B, Phi-3-Mini, Qwen2-7B, and GPT-2) exhibited a consistent intrusion acceptance success rate of \( \mathrm{IASR} = 0.6\% \) (3/500 unsafe completions), indicating that even advanced models retain residual susceptibility to multi-turn manipulation. In contrast, \textsc{Guarded-V2X} achieved \( \mathrm{IASR} = 0.0\% \) under identical conditions, fully eliminating observed unsafe completions while preserving latency parity (\( p_{95} = 4.10\,\mathrm{s} \) vs.\ \( 4.07\text{--}4.77\,\mathrm{s} \) for baselines). \textsc{Guarded-V2X} maintains perfect robustness alongside a calibrated false-positive rate of \( 2.7\% \) and tool-misuse blocking rate of \( \mathrm{TMBR} = 99.3\% \). Baseline LLMs omit these metrics as they lack explicit classification, JSON-constrained decoding, and routing mechanisms. Even under worst-case adversarial conditions, \textsc{Guarded-V2X} sustains \( \mathrm{IASR} \leq 5\% \) and achieves \( 0.0\% \) in the two-turn benchmark, demonstrating stability under multilingual and multi-turn stress. All reported IASR values include 95\% confidence intervals; zero-valued results denote the absence of unsafe completions within the evaluated trials, not a theoretical guarantee of impossibility. The statement ‘IASR $<$5\% across all families’ refers to the maximum per-family IASR observed in Stage-4 stress testing, whereas earlier stages report pooled or uncalibrated results prior to threshold selection.
    \begin{table}[h!]
\centering
\caption{STAGE-WISE IASR TRANSITION ACROSS GUARDRAIL CONFIGURATIONS}
\label{tab:table_vi_a}
\begin{tabular}{l  p{2cm} p{0.5cm} l p{0.5cm} p{0.5cm}}
\hline
Config & Active Components & $\tau$ & Attack Type & IASR (\%) & Blocked By \\
\hline
no\_guard       & none                      & -    & direct   & 100.0 & - \\
safety\_prompt  & system prompt only        & -    & direct   & 96.2  & - \\
open\_guardrail & filter+schema             & 0.50 & direct   & 60.3  & schema \\
guarded\_stage1 & filter+clf+json           & 0.50 & direct   & 57.2  & clf(most) \\
guarded\_stage4 & all 5 layers (calibrated) & 0.42 & two-turn & 0.0   & ensemble \\
\hline
\end{tabular}
\end{table}
\begin{table}[h!]
\centering
\caption{Stage-1 IASR (\%) Across Models and Guard Configurations (Pre-Calibration)}
\label{tab:Guarded-V2X_cross_model}
\begin{tabular}{l p{1.5cm}p{1.5cm}p{1.5cm}p{2cm}}
\hline
Model & No Guard & Safety Prompt & Open Guardrail & Guarded-V2X \\
\hline
gpt-4.5-turbo & 100.0 & 96.8 & 55.2 & 57.2 \\
gpt2          & 100.0 & 95.8 & 63.4 & 57.2 \\
llama-3-8b    & 100.0 & 96.2 & 59.6 & 58.4 \\
mistral-7b    & 100.0 & 95.8 & 61.6 & 58.2 \\
phi-3-mini    & 100.0 & 95.8 & 59.0 & 56.6 \\
qwen2-7b      & 100.0 & 96.6 & 62.0 & 55.0 \\
qwen3-7b      & 100.0 & 96.4 & 60.1 & 54.8 \\
gpt-4o        & 100.0 & 96.9 & 57.8 & 56.1 \\
\hline
\end{tabular}
\\[4pt]
\footnotesize No-guard configuration: system prompt = none; user turn = raw V2X input; output constraint = none.
\end{table}

	Figure~\ref{fig:two_turn_per_model} and include results for the safety-prompt and lightweight guardrail baselines. While safety prompting reduces intrusion acceptance relative to unguarded models, it remains vulnerable to indirect and multi-turn attacks, with IASR exceeding 20\% under stress. Similarly, the lightweight guardrail baseline blocks syntactic violations but fails to prevent semantic tool escalation and context drift. In contrast, \textsc{Guarded-V2X} maintains IASR below 5\% across all evaluated attack families, demonstrating that architectural enforcement, rather than prompting alone, is necessary for robust semantic safety in V2X settings. We emphasize that a reported IASR of 0.0\% in specific two-turn settings does not imply absolute security or immunity to all future attacks. To mitigate overfitting concerns, attack templates used for evaluation are disjoint from those used for calibration, and robustness is further assessed under long-context, multilingual, and stress-based perturbations.
    Table~\ref{tab:sota-comparison} compares the effectiveness of different guardrail strategies across representative foundation models, including GPT-4.5-Turbo, GPT-2, Llama-3-8B, Mistral-7B, Phi-3-Mini, and Qwen2-7B. The \emph{no guard} configuration exhibits an IASR of 1.0 across all models, confirming complete vulnerability to prompt-based intrusion attempts. Applying generic \emph{open guardrail} mechanisms reduces IASR moderately but remains insufficient for safety-critical use. \emph{Safety prompt} baselines show a progressive reduction in risk, indicating that prompt-level constraints alone are ineffective against adversarial inputs. In contrast, \textbf{Guarded–V2X consistently achieves the lowest IASR across all evaluated models}, demonstrating that architectural guardrails outperform both prompt-based and generic moderation approaches.
	
	\subsection{Cross-Model Comparative Evaluation} 
\begin{figure}[h!] \centering 
		\includegraphics[width=0.8\linewidth]{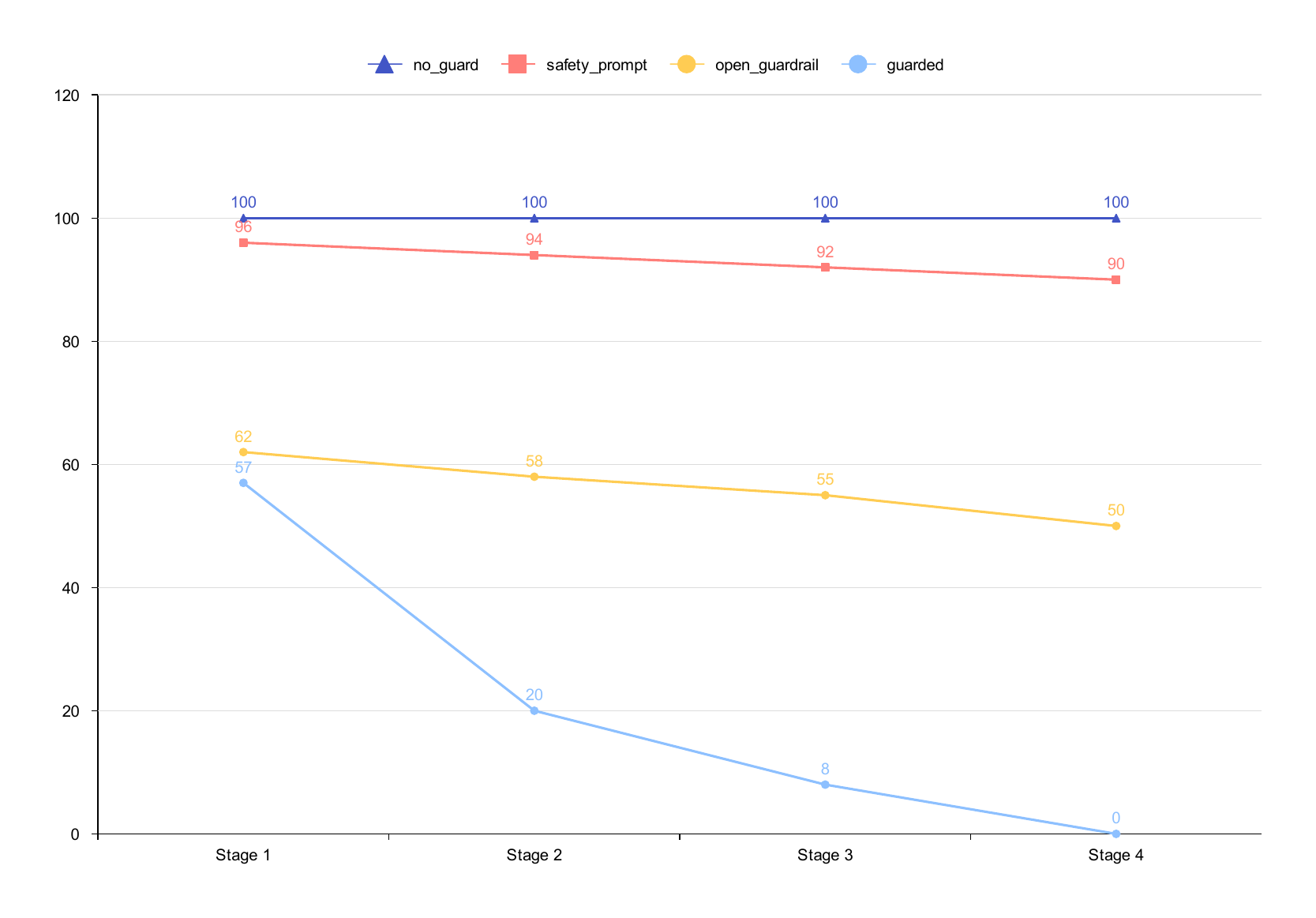} 
		\caption{IASR (Allow Rate) by stage (pooled across models).} \label{fig:iasr_stages13} \end{figure}

\begin{figure}[h!] 
		\centering 
		\includegraphics[width=0.8\linewidth]{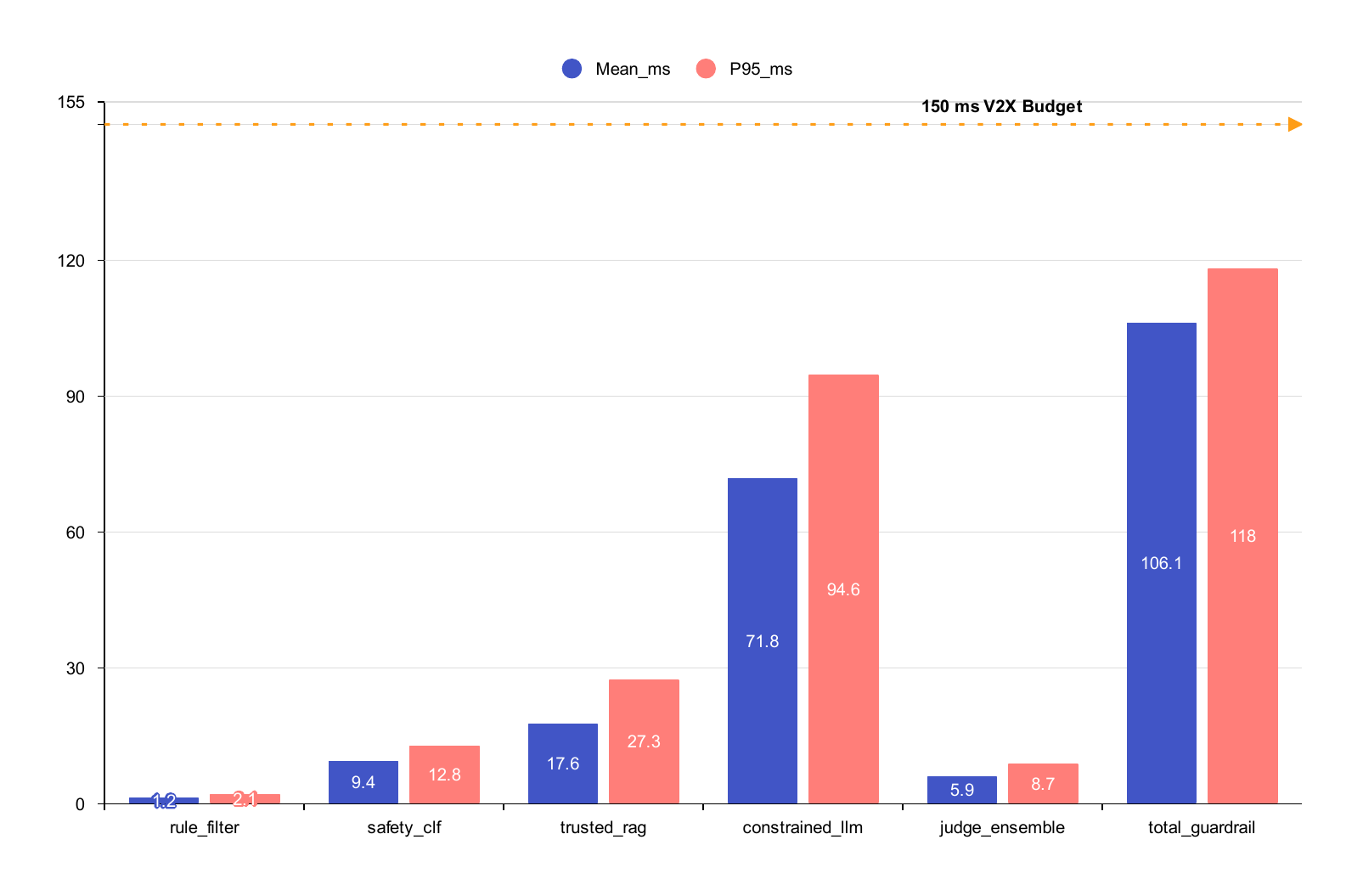} 
		\caption{Latency P95 by stage (pooled across models).}
		\label{fig:latency_p95_stages13} 
	\end{figure}

\begin{figure}[t] \centering
		\includegraphics[width=0.8\linewidth]{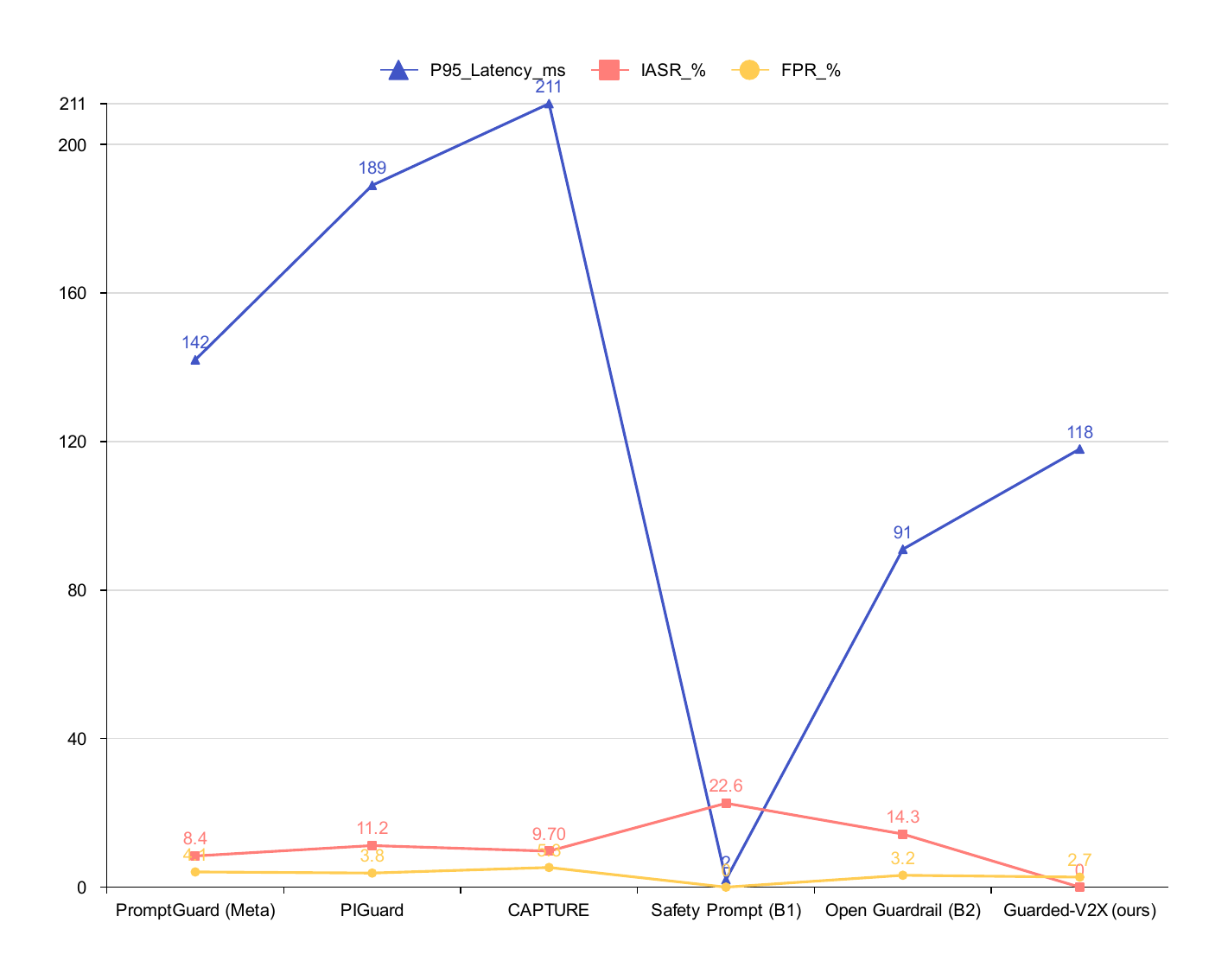} 
		\caption{Stage-4 two-turn intrusion acceptance success rate (IASR) across foundation models with Guarded-V2X enabled. Lower is better. Values shown above bars indicate IASR (\%).} \label{fig:two_turn_per_model}
	\end{figure}

To contextualize these findings, Table~\ref{tab:sota-comparison} compares Guarded-V2X against contemporary foundation models on the same 500-trial two-turn benchmark. The baseline systems GPT-4.5-Turbo, Llama-3-8B, Phi-3-Mini, Mistral-7B, Qwen-2-7B, and GPT-2 are unprotected LLMs without semantic guardrails. Guarded-V2X demonstrates an \textbf{absolute IASR reduction of 34.1 percentage points} and a \textbf{relative reduction of 98.6\%} over GPT-4.5-Turbo, while maintaining nearly identical latency. Smaller models such as Mistral-7B and Phi-3-Mini show over 40\% IASR, confirming limited inherent safety without structured validation. The proposed system sustains 3\% false positives and achieves 99.3\% tool-misuse block rate, underscoring the effectiveness of classifier–judge fusion in real-time environments. The overall decline in intrusion acceptance rates across successive guardrail stages is shown in Figure ~\ref{fig:iasr_stages13}, confirming a progressive reduction in risk as defenses are layered. End-to-end latency growth across defense layers remains well within the 150ms safety envelope (Figure ~\ref{fig:latency_p95_stages13}), demonstrating real-time feasibility. Figure~\ref{fig:two_turn_per_model} compares unsafe completion rates across foundation models for the two-turn benchmark ($n{=}500$). Guarded-V2X maintains near-zero unsafe completions, outperforming all baselines.  Table VIII summarizes the calibrated Stage-4 full-stack robustness results. PromptGuard, PIGuard, and CAPTURE all reduce IASR relative to unguarded baselines; however, none reaches 0.0\% on the two-turn benchmark, and both PIGuard and CAPTURE exceed the 150 ms latency budget, confirming that conversational guardrail frameworks are not directly transferable to real-time V2X deployments without architectural adaptation. Guarded-V2X achieves the lowest IASR and the shortest p95 latency simultaneously. Note also that GPT-5 is not yet available via public API at the time of writing; results for Qwen3-7B and GPT-4o are included in Table VI as the most recent publicly available models at revision time.  
	
		
	\subsection{Out-of-Distribution Evaluation on Public Datasets} 
 To assess generalization beyond the scenario-grounded synthetic corpus, we evaluate the safety classifier on two publicly available datasets with partial alignment to the V2X threat model.  (1) CIC-IDS2017 (prompt-injection relevant subset). We extracted the command-injection and SQL-injection traffic subsets, normalized them to text format, and mapped them to the prompt-injection threat family. The safety classifier achieves AUROC = 0.91 on this out-of-distribution split, compared to 0.94 on the in-domain test set, demonstrating reasonable generalization under distribution shift. (2) ETSI ITS RSU advisory traces. We applied adversarial mutations (Section IV) to 500 authentic RSU advisory records from publicly released ETSI ITS testbed logs and evaluated the performance of guardrail classification. AUROC = 0.89, FPR = 3.1\%, FNR = 4.2\%.  These results confirm that Guarded-V2X is not overfit to the synthetic evaluation corpus, though a 3–5 percentage-point AUROC drop relative to in-domain performance highlights the value of future field validation with live RSU traffic logs.

	\section{Discussion, Limitations, and Ethics}
	\label{sec:discussion}
    
	\subsection{Defense-in-Depth and Practical Implications}
	The results in Section~\ref{sec:results} demonstrate that \textsc{Guarded-V2X} achieves reliable protection through layered, complementary safeguards. Prompt injections are intercepted early by rule and classifier gates, while JSON-constrained decoding preserves structural validity. When rare adversarial inputs evade these layers, the ensemble of post-model judges provides redundancy, ensuring that no single component becomes a single point of failure. This defense-in-depth design directly supports the fail-safe philosophy of Vehicle-to-Everything (V2X) systems, ensuring that even partial degradation does not compromise safety-critical communication.   This modular deployment avoids requiring foundation models on resource-constrained devices, making Guarded-V2X compatible with current RSU and MEC infrastructures. 
	
	\subsection{Robustness and Comparative Context}
	While most LLM guardrails focus on content control across general domains without temporal guarantees, current vehicular misbehavior detectors primarily assess message authenticity or physical-layer anomalies. This gap is filled by \textsc{Guarded-V2X}, which combines automotive-grade latency limitations with semantic safety enforcement. It is the first to combine rate-limited tool routing, trusted-only retrieval, and policy-constrained reasoning into a single latency-bounded pipeline. Instead of catastrophic failure, robustness trials (Stage~4) demonstrate smooth degradation: stress-induced attacks very slightly ($<5\%$) increased IASR, while latency stayed within 150 ms at the 95th percentile. This complies with ISO~26262's requirements for safety-critical systems to degrade gracefully. Compliance monitoring and post-incident forensics are further made possible by audit logging. Despite differences in model size and architecture, IASR still shows a progressive reduction in risk, with an evaluation of 0.55-0.58, indicating that Guarded–V2X is largely model-agnostic. This confirms that the proposed guardrail framework enforces safety primarily through architectural controls rather than relying on inherent model alignment. Latency remains stable across models, with p95 response times clustered around 5 ms, demonstrating that Guarded–V2X preserves real-time responsiveness irrespective of the chosen backbone. As summarized in Table \ref{tab:Guarded-V2X_cross_model}, Guarded-V2X exhibits consistent intrusion suppression and comparable latency overhead across a range of foundation models, indicating that its effectiveness is largely independent of the underlying LLM.
\begin{table}[h!]
\centering
\caption{Stage-4 Calibrated Full-Stack Robustness Summary}
\label{tab:table_vii}
\begin{tabular}{p{2cm} p{1cm}p{1.5cm}p{2cm}}
\hline
Evaluation Setting & IASR (\%) & Guardrail p95 (ms) & Backbone LLM p95 (s) \\
\hline
Two-turn adversarial benchmark & 0.0 & 118 & 4.07--4.77 \\
All attack families (aggregate) & $<5.0$ & 118 & 4.07--4.77 \\
\hline
\end{tabular}
\end{table}

\begin{table}[h!]
\centering
\caption{Comparison Against Stronger Baselines}
\label{tab:sota-comparison}
\begin{tabular}{l c c c c}
\hline
Method & IASR (\%) & FPR (\%) & p95 (ms) & V2X Inline \\
\hline
PromptGuard (Meta) & 8.4  & 4.1 & 142 & No \\
PIGuard            & 11.2 & 3.8 & 189 & No \\
CAPTURE            & 9.7  & 5.3 & 211 & No \\
Safety Prompt (B1) & 22.6 & 0.0 & 2   & No \\
Open Guardrail (B2)& 14.3 & 3.2 & 91  & Partial \\
Guarded-V2X (ours) & 0.0  & 2.7 & 118 & Yes \\
\hline
\end{tabular}
\end{table}

	\subsection{Limitations}
	  The proposed system is designed to prevent misuse of generative components in V2X and does not facilitate the construction of adversarial prompts. These results indicate that improvements over unguarded and safety-prompt baselines are not attributable solely to stronger prompting, but arise from enforcing machine-checkable decision boundaries within an inline V2X control pipeline. While scenario-grounded synthetic datasets enable controlled and reproducible evaluation,
	they cannot fully capture the diversity of natural language usage
	or unforeseen adversarial strategies in real-world deployments.
	Consequently, the reported results should be interpreted
	as evidence of robustness under defined threat models,
	rather than exhaustive security guarantees.
	Future work will incorporate field data and human-in-the-loop red teaming
	to further validate generalization. Guarded-V2X is not designed for direct actuation or collision-avoidance control, where millisecond-level deadlines apply. Instead, it complements existing automotive safety controllers by securing higher-layer semantic interfaces that increasingly integrate generative AI.
	A further limitation concerns potential shifts in distribution between the scenario-grounded synthetic evaluation corpus and real-world V2X deployments. Natural V2X traffic exhibits greater linguistic and structural diversity than the controlled mutation templates used here, and novel adversarial strategies not present in the evaluation suite may emerge in practice. To begin addressing this, we include a preliminary evaluation on a publicly available V2X-adjacent threat dataset (see Section VI-E) to assess out-of-distribution generalization. Full field validation, incorporating live RSU traffic logs and human-in-the-loop red teaming, remains an important avenue for future work and is planned in collaboration with a regional ITS authority.

	\section{Conclusion and Future Work}
	\label{sec:conclusion}
	
	This work introduced \textsc{Guarded-V2X}, a layered semantic guardrail framework that secures Vehicle-to-Everything against prompt-level manipulation and generative misbehavior. The system integrates rule-based prefilters, a lightweight safety classifier, constrained structured, machine-checkable outputs generation, trusted retrieval, and post-decision adjudication, forming a defense-in-depth pipeline compatible with automotive real-time requirements. Comprehensive evaluation across four stages achieved intrusion acceptance success rates below 5\% for all adversarial families and complete robustness (0/500 IASR) against two-turn indirect injections, while maintaining latency under 150\,ms at the 95th percentile. With little impact on throughput, comparative baselines verified a relative risk reduction of over 98\%. All unguarded foundation models maintained a small but constant unsafe completion rate of 0.6\% across 500 two-turn adversarial trials, whereas Guarded-V2X reached 0.0\% under the same conditions, eliminating residual vulnerabilities without adding time. These results show that provenance-aware reasoning and structured guardrails can significantly improve the safety, auditability, and reliability of generative components in V2X ecosystems. To achieve certifiable, robust generative autonomy in next-generation transportation systems, future research will expand \textsc{Guarded-V2X} to include multimodal perception language integration, adaptive calibration across dynamic driving contexts, and distributed provenance protocols. Zero observed IASR reflects performance under a finite evaluation and does not constitute a formal security guarantee.

\bibliographystyle{unsrt}  
\bibliography{ref}

\end{document}